\documentclass{emulateapj}

\usepackage{natbib}
\def\farcs{\hbox{$.\!\!^{\prime\prime}$}}
\def\degr{\hbox{$^\circ$}}\def\arcmin{\hbox{$^\prime$}}
\def\arcsec{\hbox{$^{\prime\prime}$}}
\newcommand{\cii}{[\ion{C}{2}]}
\newcommand{\ho}{H$_2$O}
\newcommand{\kms}{km\,s$^{-1}$}
\newcommand{\lam}{$\lambda$}
\newcommand{\um}{$\rm \mu$m}
\newcommand{\ergs}{erg\,s$^{-1}$}

\newcommand{\wm}{W\,m$^{-2}$}
\newcommand{\cmc}{cm$^{-3}$}

\newcommand{\msol}{M$_{\odot}$} 
\newcommand{\lsol}{L$_{\odot}$} 
\newcommand{\mearth}{M$_{\earth}$} 
\newcommand{\be}{\begin{equation}} 
\newcommand{\ee}{\end{equation}} 

\begin{document}

\title{Water vapor in the protoplanetary disk of DG Tau}

\author{L. Podio\altaffilmark{1}, I. Kamp\altaffilmark{2}, C. Codella\altaffilmark{3}, S. Cabrit\altaffilmark{4}, B. Nisini\altaffilmark{5}, C. Dougados\altaffilmark{1}, G. Sandell\altaffilmark{6}, J.~P. Williams\altaffilmark{7}, L. Testi\altaffilmark{8}, W.-F. Thi\altaffilmark{1}, P. Woitke\altaffilmark{9}, R. Meijerink\altaffilmark{2}, M. Spaans\altaffilmark{2}, G. Aresu\altaffilmark{2}, F. Menard\altaffilmark{1, 10}, C. Pinte\altaffilmark{1}}

\altaffiltext{1}{UJF-Grenoble 1 / CNRS-INSU, Institut de Plan\'etologie et d'Astrophysique de Grenoble (IPAG) UMR 5274, Grenoble, F-38041, France}
\altaffiltext{2}{Kapteyn Astronomical Institute, University of Groningen, Landleven 12, 9747 AD Groningen, The Netherlands} 
\altaffiltext{3}{INAF - Osservatorio Astrofisico di Arcetri, Largo E. Fermi 5, 50125, Florence, Italy} 
\altaffiltext{4}{LERMA, UMR 8112 du CNRS, Observatoire de Paris, \'Ecole Normale Sup\'erieure, Universit\'e Pierre et Marie Curie, Universit\'e de Cergy-Pontoise, 61 Av. de l'Observatoire, 75014, Paris, France}
\altaffiltext{5}{INAF - Osservatorio Astronomico di Roma, via di Frascati 33, 00040, Monte Porzio Catone, Italy}  
\altaffiltext{6}{SOFIA-USRA, NASA Ames Research Center, MS 232-12, Building N232, Rm. 146, P. O. Box 1, Moffett Field, CA 94035- 0001, U. S. A.}
\altaffiltext{7}{Institute for Astronomy (IfA), University of Hawaii, 2680 Woodlawn Dr., Honolulu, HI, 96822, USA}
\altaffiltext{8}{European Southern Observatory, Karl-Schwarzschild-Strasse 2, D-85748 Garching, Germany}  
\altaffiltext{9}{SUPA, School of Physics and Astronomy, University of St. Andrews, KY16 9SS, UK}  
\altaffiltext{10}{LFCA , UMI 3386, CNRS and  Dept. de Astronomia, Universidad de Chile, Santiago, Chile}

\begin{abstract}
Water is key in the evolution of protoplanetary disks and the formation of comets and icy/water planets. 
While high excitation water lines originating in the hot inner disk have been detected in several T Tauri stars (TTSs), water vapor from the outer disk, where most of water ice reservoir is stored, was only reported in the closeby TTS TW~Hya.
We present spectrally resolved {\it Herschel}/HIFI observations of the young TTS DG~Tau in the ortho- and para- water ground-state transitions at 557, 1113 GHz. The lines show a narrow double-peaked profile, consistent with an origin in the outer disk, and are $\sim19-26$ times brighter than in TW Hya.
In contrast, CO and \cii\ lines are dominated by emission from the envelope/outflow, which makes \ho\ lines a unique tracer of the disk of DG Tau.
Disk modeling with the thermo-chemical code ProDiMo indicates that the strong UV field, due to the young age and strong accretion of DG~Tau, irradiates  a disk upper layer at 10--90~AU from the star, heating it up to temperatures of 600 K and producing the observed bright water lines. 
The models suggest a disk mass of 0.015--0.1~\msol, consistent with the estimated minimum mass of the solar nebula before planet formation, and a water reservoir of $\sim10^2-10^3$ Earth oceans in vapour, and $\sim100$ times larger in the form of ice.    
Hence, this detection supports the scenario of ocean delivery on terrestrial planets by impact of icy bodies forming in the outer disk.
\end{abstract}


\section{Introduction}

Protoplanetary disks are the birthplaces of planets, thus the study of  their physical and chemical structure is fundamental  to comprehend the formation of our own solar system as well as of extra-solar planetary systems. 
One of the most intriguing issues on planet formation concerns the origin of oceans on Earth. It was argued that Earth formed "dry" and that ocean water was delivered by impacts of icy bodies/protocomets originating from the cold outer disk, where most of the mass (and water reservoir) is located \citep{matsui86}. To address this issue, many efforts have been devoted to observe water in protoplanetary disks and to characterize its abundance and spatial distribution. 

In the hot dense inner disk region inside the so-called 'snow line' where $T_{\rm dust}\sim150$~K, i.e. for radii smaller than $\sim1-3$~AU in disks around T Tauri Stars (TTSs) \citep{lecar06}, ice cannot exist on dust grains and gas-phase chemistry converts all oxygen into water on timescales short compared to the disk evolution timescale. Beyond the snow-line, instead, water molecules will be frozen onto dust grains. However, (inter)stellar UV and X-ray radiation can penetrate in the disk upper layers and photo-desorb a fraction of water ice back into the gas phase \citep{ceccarelli05,dominik05}. The released water vapour may be eventually dissociated and re-formed in the gas-phase.  

\ho\, lines with upper level energies $E_{\rm up}>1000$~K, tracing hot water vapor in the inner disk regions, have now been observed in a number of protoplanetary disks thanks to ground-based and Spitzer near- and mid-infrared observations \citep[e.g. ][]{carr08,salyk08,pontoppidan10b,pontoppidan10a}, and, recently, far-infrared observations of the 63.32~\um\ line with {\it Herschel} \citep{riviere-marichalar12}. 
In contrast, cold water vapor at $T<200$~K from the outer disk surface  has turned out surprisingly difficult to detect in TTSs. {\it Herschel}/PACS detected the low excitation \ho~179.5~\um\ line ($E_{\rm up}=114$~K) only in jet-driving stars but due to the lack of spatial and velocity information, it is unclear if it originates in the disk or in the envelope/outflow \citep{podio12}. 
Up to now, firm evidence for a cold disk water reservoir has been found only in the nearby ($d\sim50$~pc) TTS TW Hya, through the detection of the fundamental ortho and para lines at 557 and 1113 GHz  with {\it Herschel}/HIFI \citep{hogerheijde11}. 
While o-\ho~557~GHz line profiles in Class 0 and I sources show velocities of $\sim11-138$~\kms, and $\sim5-54$~\kms, suggesting they are dominated by emission from the envelope/outflow \citep{kristensen12}, the \ho\ emission from TW Hya shows a narrow single-peaked profile ($FWHM\sim0.96-1.2$~\kms) consistent with an origin in the face-on disk.
A hidden reservoir of icy bodies of 1.5~\mearth, equivalent to several thousands of Earth oceans\footnote{1 \mearth$=5.97\,10^{27}$ g, 1 Earth ocean $\simeq1.5\,10^{24}$ g} is inferred. Additional studies are necessary to investigate this hypothesis, but only upper limits were obtained toward a couple of other TTSs targeted with HIFI, e.g. DM Tau \citep{bergin10}.  

DG Tau is a young TTS at 140~pc associated with particularly strong accretion/outflow activity \citep[e.g. ][]{hartigan95,dougados00}, and where we previously detected unresolved emission in the \ho~78.7, 179.5~\um\ lines with {\it Herschel}/PACS \citep{podio12}. In this Letter, we present clear detections of the \ho~557, 1113 GHz lines towards this source. 


\section{Observations and data reduction}
\label{sect:obs}


We observed DG Tau ($\alpha_{\rm J2000}=04^{\rm h}$ 27$^{\rm m}$ 04$\fs$7, $\delta_{\rm J2000}=+26\degr$ 06$\arcmin$ 16$\farcs$3) with the Heterodyne Instrument for the Far Infrared (HIFI, \citealt{degraauw10}) on board the {\it Herschel}  Space Observatory\footnote{{\it Herschel} is an ESA space observatory with science instruments provided by European-led Principal Investigator consortia and with important participation from NASA.} \citep{pilbratt10}. 
The observations target the two fundamental water lines, o-\ho~1$_{10}$-1$_{01}$ and p-\ho~1$_{11}$-0$_{00}$, the $^{12}$CO (hereafter CO) and $^{13}$CO 10--9, and the \cii~$^2$P$_{3/2}$-$^2$P$_{1/2}$ lines (OBSID: 1342239630, 1342250208, 1342249594, 1342249646). 
They were acquired in the HIFI bands 1, 4, 5, and 7, with a single on-source pointing and in dual beam switch mode with fast chopping 3\arcmin\ either side of the target. The Wide Band Spectrometer (WBS) and the High Resolution Spectrometer (HRS) were used in parallel, with a spectral resolution of 1.10 and 0.25~MHz, respectively. The Half Power Beam Width (HPBW) ranges from $\sim11$\arcsec\ to $\sim38$\arcsec, depending on frequency.

HIFI data were reduced using HIPE 8\footnote{HIPE is a joint development by the {\it Herschel} Science Ground Segment Consortium, consisting of ESA, the NASA {\it Herschel} Science Center, and the HIFI, PACS and SPIRE consortia.}.
Fits files from level 2 were then created and transformed into GILDAS\footnote{http://www.iram.fr/IRAMFR/GILDAS} format for data analysis. The spectra were baseline subtracted and then resampled at 0.6~\kms\ to increase the sensitivity.
Note that the V-spectrum of the o-\ho\ and p-\ho\ lines are affected by ripples, degrading the quality of the baseline and resulting in an rms larger than the one measured in the H-spectrum. Therefore, in the following, we will analyse the o-\ho\ and p-\ho\ emission based solely on the H-spectrum.

The HIFI dataset is complemented by observations of the CO 3--2 line performed  on January 2010 at the JCMT 15m telescope (Manua Kea, Hawaii, USA) using the HARP-B heterodyne array and ASCIS correlator, providing a spectral resolution of 0.25~\kms. The spectrum was resampled at 0.6~\kms~ to be compared with the HIFI data.

Antenna temperatures, $T_{\rm a}$, are converted to mean beam temperature, $T_{\rm mb}$ (for HIFI mean beam efficiency are by \citealt{roelfsema12}). 
Integrated line intensities, $\int T_{\rm mb}dV$, and line fluxes, $F_{\rm obs}=\frac{2K_{\rm b}\nu^3}{c^3}\times\int T_{\rm mb}dV\times\pi\left(\frac{HPBW}{2\sqrt{\ln2}}\right)^2$, are summarized in tables \ref{tab:lines_obs} and \ref{tab:obs_model}.

\section{Results from observations}
\label{sect:results}

\begin{table}
\caption{\label{tab:lines_obs}Lines integrated intensities}
\begin{tabular}{ccccc}
\tableline\tableline
Transition$^a$ & 
$\nu_{\rm 0}$$^b$ & 
$\eta_{\rm mb}$  &
HPBW &
$\int T_{\rm mb}dV$ \\
 &
GHz &
&
\arcsec &
 K \kms \\
\tableline
o-\ho\ 1$_{10}$ - 1$_{01}$ & 
556.936 &
0.76 &
38 &
0.10 $\pm$ 0.01 \\
p-\ho\ 1$_{11}$ - 0$_{00}$ & 
1113.343 &
0.74 &
19 &
0.12 $\pm$ 0.01 \\
CO 10--9 & 
1151.985 &
0.64 &
18 &
5.8 $\pm$ 0.1 \\
$^{13}$CO 10--9 & 
1101.350 &
0.74 &
19 &
0.28 $\pm$ 0.01 \\
\cii\ $^2$P$_{3/2}$-$^2$P$_{1/2}$ & 
1900.537 &
0.69 &
11 &
3.1 $\pm$ 0.2 \\
CO 3--2 & 
345.795 &
0.66 &
14 &
32.5 $\pm$ 0.4 \\
\tableline
\end{tabular}
$^a$All lines are observed with {\it Herschel}/HIFI except CO 3--2 which is observed with JCMT/HARP-B\\ 
$^b$Frequencies are from the Jet Propulsion Laboratory molecular database \citep{pickett98}\\
\end{table}

\begin{figure}[!ht]
     \centering
     \includegraphics[width=8.8cm]{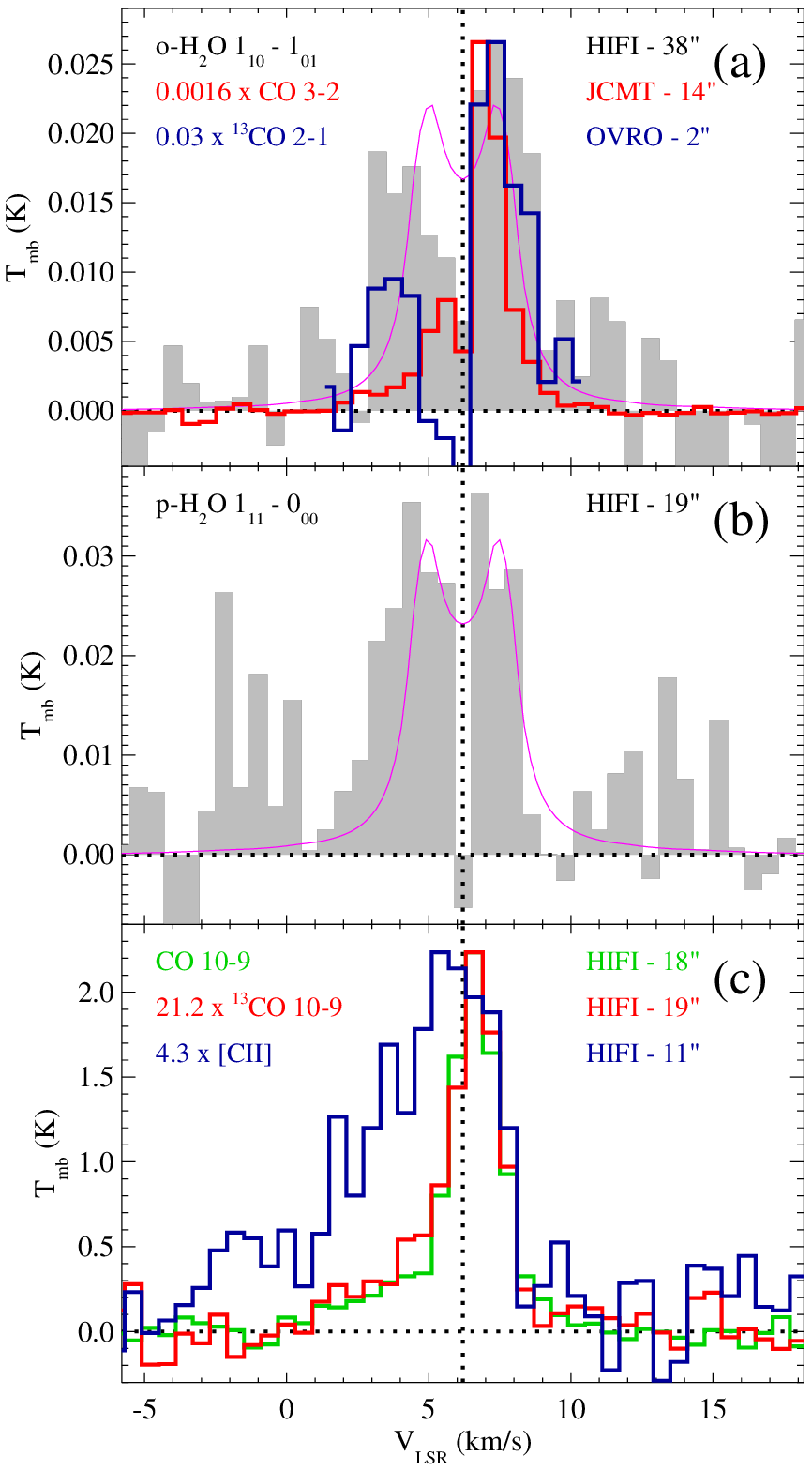}
   \caption{HIFI spectra of (a) o-\ho~1$_{10}$-1$_{01}$ (grey histogram); (b) p-\ho~1$_{11}$-0$_{00}$ (grey histogram); (c) CO 10--9, $^{13}$CO 10--9, and \cii~$^2$P$_{3/2}$-$^2$P$_{1/2}$ (green, red, and blue histograms, respectively). In panel (a) JCMT CO 3--2 (red histogram), and $^{13}$CO 2--1 profile obtained integrating interferometric maps by \citet{testi02} on a 2\arcsec\ beam (blue histogram) are also shown.  The vertical dotted line indicates the systemic velocity ($V_{\rm LSR}=+6.2$~\kms). The \ho\ line profiles predicted by the ``low dust opacity'' ProDiMo disk model are overplotted (magenta lines). The o-\ho~ line flux is underpredicted by the model by a factor $\sim$2.2, hence the line profile is multiplied by this factor to help the comparison with observations.}
   \label{fig:hifi_lines}
    \end{figure}

The observed line profiles are shown in figure \ref{fig:hifi_lines}. The JCMT CO 3--2 line profile in panel (a) suggests that the systemic velocity is $V_{\rm sys}\sim+6.2$ \kms, consistent with previous studies \citep{schuster93,kitamura96a,testi02}.  
 
We detect both the ortho-\ho~1$_{10}$-1$_{01}$~557~GHz and the para-\ho~1$_{11}$-0$_{00}$~1113~GHz lines ($E_{\rm up}\sim61$, 53~K) with a signal-to-noise of 10 and 12, respectively. They are centered at the systemic velocity and show a narrow double-peaked profile ($FWHM\sim5-6$~\kms). 

CO 10--9, $^{13}$CO 10--9, and \cii~158~\um\ lines have a different profile than \ho, with a single peak near systemic velocity (at $V_{\rm LSR}=+6.5$ \kms\ in CO and $+5.5$~\kms\ in \cii), and a pronounced blue wing extending down to 0 and $-5$~\kms\ respectively (i.e. 6 and 11~\kms\ away from systemic). 
The bulk of CO and \cii\ emission close to systemic velocity likely originates in the envelope, as suggested by the $^{13}$CO 2--1 channel maps by \citet{testi02} that indicate emission extended over $\simeq10$\arcsec\ at velocities $\mid$$V-V_{\rm sys}$$\mid<1.5$~\kms. The observed blue wing, instead, may originate in a slow outflow, perhaps linked to envelope dispersal motions as proposed by \citet{kitamura96a}.  For the \cii~158~\um\ line an origin in an extended structure is further confirmed by the fact that the flux in the HIFI beam of $\sim11$\arcsec\ is $\sim4$ times lower than the total co-added flux in the {\it Herschel}/PACS observations (47\arcsec$\times$47\arcsec, \citealt{podio12}). 

On the other hand, several arguments suggest that the \ho\ emission is compact and is likely dominated by emission from the outer region of the protoplanetary disk of DG Tau, and not from the envelope/outflow:
\begin{itemize}
\item The \ho\ line profiles are different from those of CO and \cii~ observed with single-dish telescopes. They are much more symmetric about the systemic velocity and do not show the extended blue wing seen in these other tracers.
\item The peaks of the \ho\ line profiles coincide with the two narrow velocity ranges ($\mid$$V-V_{\rm sys}$$\mid=1.5-2.5$~\kms) where $^{13}$CO 2--1 interferometric maps show compact emission with a velocity gradient perpendicular to the jet axis, consistent with disk rotation \citep{testi02}. The $^{13}$CO 2--1 line profile obtained by integrating the interferometric maps by \citet{testi02} over a 2\arcsec\ beam, i.e. by cutting out any extended component, is similar to the \ho\ line profiles, with peaks at the same velocities.
In contrast, the CO 3--2 profile, obtained with the JCMT collecting all the emission in the 14\arcsec\ beam, does not peak at the same velocity as the \ho\ and $^{13}$CO compact component. This is particularly clear in the blue part of the profile.
%
\item Assuming keplerian rotation, and an inclination of $i\simeq38$\degr\ from the line of sight \citep{eisloffel98}  the peak separation of the \ho\ lines ($\Delta V_{\rm sep}\sim3-3.5$~\kms) indicates an outer disk radius $R_{\rm out}$(\ho)$\sim77-105 (M_\star/0.7M_\odot)$~AU. For a stellar mass of $\sim0.7$~\msol, as assumed in \citet{testi02}, the inferred $R_{\rm out}$(\ho) is in agreement with the disk outer radius, $\sim72-89$~AU, estimated from sub-arcsecond dust continuum maps at 1.3, 2.8~mm with CARMA \citep{isella10}. The maximum velocities covered by the line profiles, instead, set an upper limit to the inner radius of the line emitting region $R_{\rm in}$(\ho)$\le19$~AU, since more extended line wings could be hidden in the noise.  
\item  The \ho\ line profiles are reproduced by an optically thick, vertically isothermal keplerian disk with $T_{\rm ex}\propto r^{-0.5}$  viewed at 38\degr\ with an excitation temperature at $R_{\rm out}$ of 70 and 32 K for the ortho and para lines respectively  \citep{beckwith93,cabrit06}.
\end{itemize}
Given the evidence listed above, the fundamental water lines, even when observed with a 38\arcsec--19\arcsec\ beam, appear to be dominated by compact emission. Although we cannot exclude contamination from the outflow, which could explain the larger FWHM and the asymmetry of the o-\ho~557~GHz, the detected double-peaked \ho\ lines prove to be a good tracer of the outer protoplanetary disk of DG Tau, with less confusion from envelope/outflow than in $^{13}$CO.

DG Tau shows emission also in high-excitation \ho\ lines observed with PACS \citep{podio12}. With  $E_{\rm up}\sim200-1070$~K these are thought to originate in an intermediate disk region between a few and a few tens of AU from the star \citep[e.g. ][]{riviere-marichalar12}.
The exception is the low-excitation \ho~179.5~\um\ line ($E_{\rm up}\sim114$~K) which according to previous disk modeling is predicted to form in the outer disk like the 557, 1113 GHz lines  (Kamp et al. submitted). 
The observed \ho~179.5~\um~/~557~GHz line ratio is $R1_{\rm obs}=25\pm6$, consistent with LTE optically thick emission in the Rayleigh-Jeans limit, i.e. for temperatures larger than a few hundreds K ($R_{\rm LTE-thick}\sim27$). 
On the other hand, the line ratio between the para- and the ortho- fundamental lines ($R2_{\rm obs}=2.5\pm0.3$) is around three times lower than $R_{\rm LTE-thick}\sim8$. This can be explained if the lines are excited in a region where the gas density is lower than the lines critical density ($\sim2\,10^{7}$ and $\sim2\,10^{8}$ at 50~K for the 557 and 1113 GHz lines) and/or where the temperature is below their upper level energies. Also, the observed line ratio could be affected by emission from the envelope/outflow. \\

\section{Modeling \ho\ in the disk of DG Tau}
\label{sect:modeling}

   \begin{figure}
     \centering
     \includegraphics[width=8.8cm]{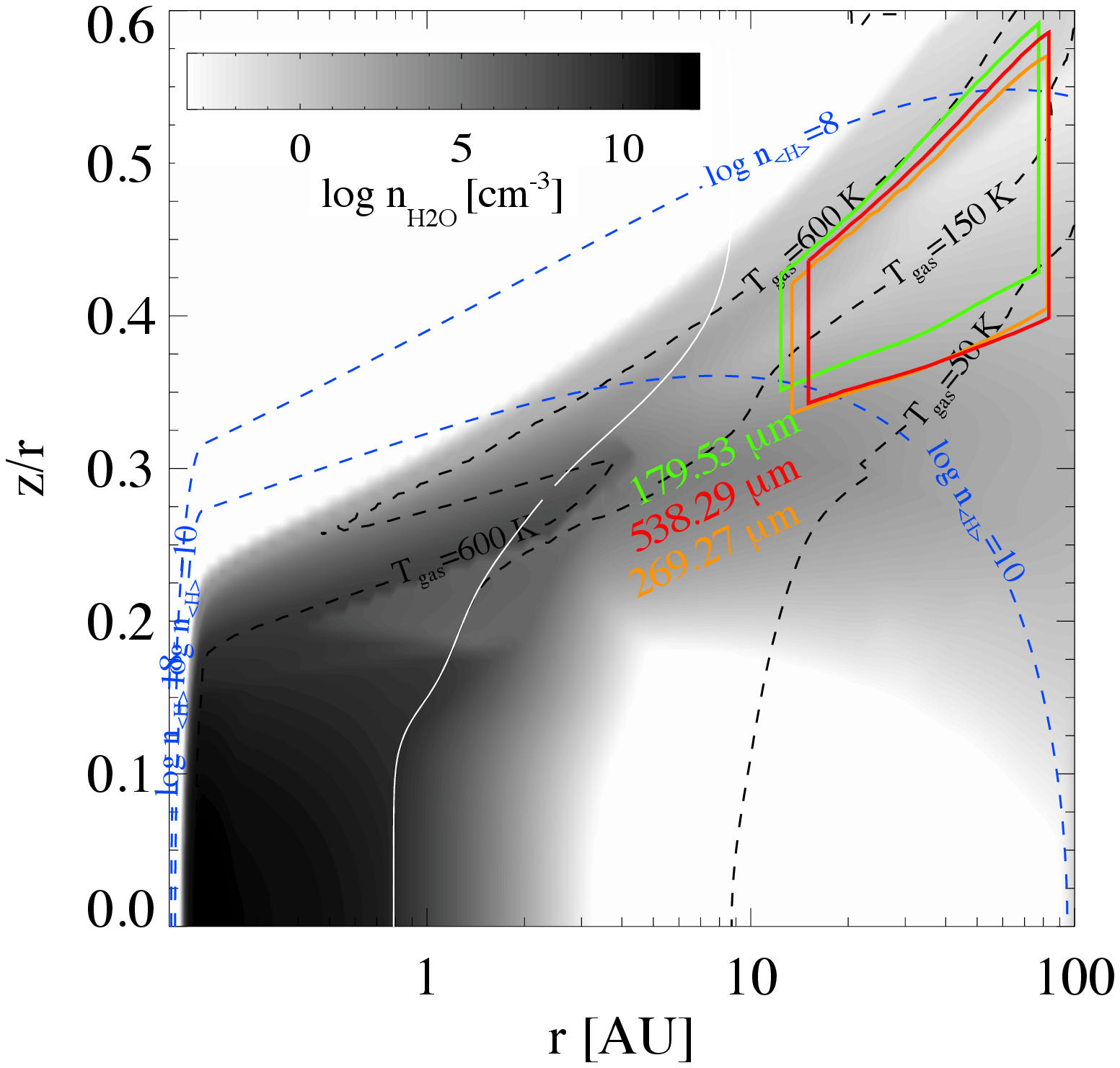}
   \caption{Disk region from which 50\% of the o-\ho~179.5~\um\ (in green), o-\ho~538.3~\um\ (or 557 GHz, in red), and p-\ho~269~\um\ (or 1113 GHz, in orange) line emission arises according to the ``low dust opacity'' disk model. The grey colour indicates the water density, $n_{\rm H_{2}O}$ (\cmc), the dotted black and blue curves the gas temperature and density, and the white solid curve the snow-line (i.e. $T_{\rm dust}=150$~K).}
   \label{fig:h2o_emitting_region}
    \end{figure}

Detailed disk modeling is required to test the disk hypothesis and to derive an estimate of the water mass. The latter cannot be inferred from observations since the lines are likely optically thick.      
We include in our analysis the fluxes and upper limits obtained for the water lines falling between 63.3 and 180.5~\um\ observed with PACS as part of the {\it Herschel} Key Project GASPS (PI: B. Dent) \citep{podio12}. The two detected o-\ho\ lines at 78.7 and 179.5~\um\ are spectrally and spatially unresolved, thus their origin is unclear. In  \citet{podio12} a shock-origin was favoured based on the large line fluxes, which are difficult to reproduce with disk models for typical TTSs parameters. However, since the profiles of the ground-state water lines are consistent with a disk origin, we test the predictions of a dedicated model for DG Tau by comparing them with observed \ho\ line fluxes and profiles.


We use a parametrized disk model calculated with the thermo-chemical disk modeling code ProDiMo \citep{woitke09,kamp10}. We adopt the stellar spectral type K7 ($T_{\rm eff}\simeq4000$~K) and veiling-corrected stellar radius of 1.8~$R_\odot$ \citep{fischer11}. The resulting stellar luminosity $\simeq1$~\lsol\ yields a stellar mass $M_{\star}\simeq0.7$~\msol\ and an age of 2.5\,10$^6$~yr using the evolutionary tracks of \citet{siess00}. 
To reproduce the IUE UV/optical spectrum \citep{gullbring00}, we set the UV excess fraction $f_{\rm UV}=L(910-2500~\AA)/L_\ast=0.2$ and adopt a power law slope $L_{\rm \lambda}\approx\lambda^{-0.3}$. 
We also account for the effect of X-ray radiation from the stellar corona ($L_{\rm X}=10^{30}$~\ergs, \citealt{gudel07a}) following \citet{aresu11,meijerink12}.
The disk inner and outer radius are set to $R_{\rm in}=0.16$~AU \citep{akeson05}, and $R_{\rm out}=100$~AU, in agreement with $R_{\rm out}$(\ho) inferred from the observed \ho\ profiles. 
We assume the dust size distribution and disk dust mass from the ``low dust opacity model'' --- a 50/50 mixture of astronomical silicates \citep{draine84} and amorphous carbon \citep{zubko96} --- used by \citet{isella10} to reproduce the observed 1.3, 2.8 mm emission ($n(a)\approx a^{-q}$ with $q=3.5$, where $a$ is the dust grain radius, the minimum/maximum grain size are $a_{\rm min}=0.005$~\um\ and $a_{\rm max}=5$~cm). Using the standard dust-to-gas ratio of $0.01$ the gas mass is set to 0.1~\msol. The disk is thought to be perpendicular to the jet, thus $i=38$\degr \citep{eisloffel98}. We assume a parametrized disk shape with surface density $\Sigma\approx r^{-1}$, and scale height $H=0.008~{\rm AU}~\left(r/0.16~{\rm AU}\right)^{1.2}$. No dust settling is invoked, i.e. dust and gas are well mixed throughout this young disk. The PAH fraction is $0.01$ with respect to the ISM abundance of $10^{-6.52}$ PAH particles/H-nucleus. 
All parameters adopted for the model are summarized in table \ref{tab:dgtau_par}.

The line profiles and fluxes are obtained by first solving the statistical equilibrium with 2D escape probability to obtain the level populations, and then using 2D radiative transfer (collision rates as listed in table 3 of Kamp et al., submitted). The region from which 50\% of the \ho\ line emission arises, instead, is obtained using vertical escape probability and without accounting for disk inclination. 
Figure \ref{fig:h2o_emitting_region} indicates that the \ho~179.5~\um\ line observed with PACS originates in the same region as the fundamental water lines at 557, 1113 GHz observed with HIFI, i.e. in an upper disk layer ($z/r\sim0.35-0.6$) located at $\sim10-90$~AU distance from the star. 
In this region the gas temperature is $\sim50-600$ K and water is formed mainly through gas-phase reactions and partially dissociated by UV photons and collisions with C$^{+}$ and H$^{+}$. Including self-shielding for all photodissociating species produces at most 20\% lower fluxes.
The gas density is $10^{8}-10^{10}$~\cmc\ thus, as suggested by the observed \ho~179.5~\um/557 GHz line ratio, these lines are close to LTE and optically thick ($\tau\sim10^3-10^4$). The ortho-to-para ratio (OPR) is calculated from the gas temperature at thermal equilibrium and is 1.5--3 in the line emitting region. However, since the \ho\ lines are optically thick, the model results are not dependent on the OPR. 

As shown in figure \ref{fig:hifi_lines} the model reproduces the p-\ho\ line flux and profile, and the ratio o-\ho~179.5~\um~/~557~GHz is $R1_{\rm mod}\simeq25$, in agreement with the observed value.
On the other hand, the observed ortho lines at 179.5~\um\ and 557~GHz are underpredicted by a factor $\sim2$.
As a consequence the observed p-\ho~1113~/~o-\ho~557 line ratio is overpredicted by a factor 2.4 ($R2_{\rm mod}=6.1$).
Kamp et al. (submitted) discuss in detail the uncertainties when modeling water emission in disks. They show that the assumed surface chemistry, adsorption energy and photo-desorption yields, and metal abundances can affect water line fluxes by a factor of a few. In particular, the low-excitation water lines are very sensitive to the adopted radiative transfer method and to the uncertainties in the collision rates. 
Moreover the disk model is not accounting for the X-ray emission by the jet \citep{gudel08} which illuminates the disk surface from above. This may boost water formation through H$_3$O$^{+}$ recombination ($H_3O^{+}+e^{-}\rightarrow H_2O+H$, \citealt{meijerink12}).
In general, the model can reproduce all the \ho\ lines observed with PACS and HIFI within a factor 2  (see table \ref{tab:obs_model}). 
The emission in the CO and \cii\ lines is predicted to be a factor 3-9 lower than observed, which is in agreement with the observed profiles indicating that the bulk of the emission originate from the envelope/outflow.    

The disk model indicates that the disk contains $\sim0.4$ \mearth\  of water vapour, and two orders of magnitude larger mass in ice: M(\ho\#) $\sim100$ \mearth.
To understand the reliability of the estimated water mass in the disk, we calculate a second model assuming the dust size distribution and disk dust mass from the ``high dust opacity model'' by \citet{isella10}.
This implies around an order of magnitude lower dust mass in the disk and consequently around an order of magnitude lower gas mass, and water vapor and ice mass ($M_{\rm gas}=0.015$ \msol, M(\ho)$\sim0.06$ \mearth,  M(\ho\#)$\sim7$ \mearth).
We find that this model can reproduce equally well the observed \ho\ line fluxes, because the ``high dust opacity model'' implies lower opacity at UV wavelengths and thus a deeper UV penetration in the outer disk regions.
Hence, the dust size distribution is crucial to constrain the disk mass and water reservoir, leading to an uncertainty of one order of magnitude. The total water reservoir,  M(\ho)$_{\rm gas+ice}\sim7-100$ \mearth, is a factor of a few up to two orders of magnitude larger than for TW Hya \citep{hogerheijde11}.

\begin{table}
\caption{\label{tab:dgtau_par}``low dust opacity'' disk model: star and disk parameters}
    \begin{tabular}{lll}
\tableline\tableline
Effective temperature & $T_{\rm eff}$ (K)        & 4000  \\
Stellar mass                & $M_{*}$ (\msol)       & 0.7    \\       
Stellar luminosity       & $L_{*}$ (\lsol)          & 1       \\    
UV excess                   & $f_{\rm UV}$              & 0.2    \\
UV power law index    & $p_{\rm UV}$             & -0.3   \\             
X-rays luminosity       & $L_{\rm X}$ (\ergs)    & 10$^{30}$ \\ 
Disk inner radius        & $R_{\rm in}$ (AU)        & 0.16   \\
Disk outer radius        & $R_{\rm out}$ (AU)       & 100    \\
Disk dust mass           & $M_{\rm dust}$ (\msol) & 1 10$^{-3}$ \\
Dust-to-gas ratio       &  dust-to-gas             & 0.01 \\ 
Solid material mass density & $\rho_{\rm dust}$ (g~\cmc) & 3.5 \\
Minimum grain size    & $a_{\rm min}$ (\um)    & 0.005 \\
Maximum grain size   & $a_{\rm max}$ (cm)    & 5  \\
Dust size distribution index & $q$              & 3.5 \\
Disk inclination          & $i$ (\degr)              & 38 \\
Surface density $\Sigma\approx r^{-\epsilon}$ & $\epsilon$ & -1 \\
Scale height at $R_{\rm in}$ & $H_{\rm 0}$ (AU)         & 0.008 \\
Disk flaring index $H(r)=H_{\rm 0}\left(\frac{r}{R_{\rm in}}\right)^{\beta}$  & $\beta$ & 1.2 \\ 
Fraction of PAHs w.r.t. ISM & $f_{\rm PAH}$      & 0.01 \\
\tableline
    \end{tabular}
\end{table}

\begin{table}
\begin{center}
\caption{\label{tab:obs_model} Observed and disk-model predicted \ho\ fluxes}  
    \begin{tabular}{ccccc}
\tableline\tableline
Line & \lam &  $E_{\rm up}$ & $F_{\rm obs} \pm \Delta F$ & $F_{\rm mod}$ \\ 
        & \um &  K & \wm  & \wm \\ 
\tableline 
\multicolumn{5}{c}{PACS Observations} \\
\tableline 
    o-\ho     &  63.3     & 1070 &            $\le$          4  10$^{-17}$  & 1.9 10$^{-17}$ \\ 
    o-\ho     &  71.9     & 843  &            $\le$          1  10$^{-17}$  & 1.7 10$^{-17}$ \\ 
    o-\ho     &  78.7     & 432  &        1.9 $\pm$        1.4  10$^{-17}$ &  2.1 10$^{-17}$ \\ 
   o-\ho     & 179.5     & 114  &        1.5 $\pm$        0.3  10$^{-17}$ & 7.7 10$^{-18}$ \\ 
  o-\ho     & 180.5     & 194  &            $\le$          1  10$^{-17}$ & 3.2 10$^{-18}$ \\  
   p-\ho     &  78.9     & 781  &            $\le$          1  10$^{-17}$  & 9.7 10$^{-18}$ \\ 
    p-\ho     &  89.9     & 297  &            $\le$          1  10$^{-17}$  & 1.4 10$^{-17}$ \\ 
   p-\ho     & 144.5     & 396  &            $\le$          1  10$^{-17}$ & 2.7 10$^{-18}$ \\ 
   p-\ho     & 158.3     & 410  &            $\le$          1  10$^{-17}$  & 3.9 10$^{-19}$ \\ 
\tableline 
\multicolumn{5}{c}{HIFI Observations} \\
\tableline 
 o-\ho     & 538.3 & 61 & 6.7 $\pm$ 0.7 10$^{-19}$ & 3.1 10$^{-19}$ \\ 
 p-\ho     & 269.3 & 53 & 1.7 $\pm$ 0.2 10$^{-18}$ & 1.9 10$^{-18}$ \\ 
\tableline 
    \end{tabular}
\end{center}
\end{table}

\section{Conclusions}
\label{sect:conclusions}


The present detection of the o-\ho\ and p-\ho\ lines at 557, 1113 GHz in the TTS DG Tau is crucial for several reasons: (i) so far, emission in the fundamental water lines has been observed only in one TTS, TW Hya; (ii) we detect for the first time a double-peaked profile in the \ho\ lines, which is a strong kinematic evidence for an origin in the outer disk (from $\sim10-90$ AU); (iii) water turns out to be a unique tracer of the protoplanetary disk of DG Tau, because is less contaminated by envelope/outflow emission than CO lines; (iv) once corrected for distance the \ho\ lines are $\sim19-26$ times brighter than in TW Hya. According to our models, the
reason is the 10 times higher UV flux of DG Tau, which heats the outer disk surface layer up to temperatures of $\sim 600$~K (only $\sim 30$~K in the case of TW Hya). In addition, the disk around DG Tau is more massive and compact leading to higher volume densities in the surface layers, which makes the warm neutral chemistry even more efficient; (v)  the adopted models suggest a disk mass of  0.015--0.1 \msol, depending on the assumed dust size distribution, and a water reservoir (gas+ice) of 7--100 \mearth, i.e. at least a factor of a few larger than estimated for TW Hya \citep{hogerheijde11}.

While the inferred disk mass is consistent with the minimum mass of the solar nebula to form our solar system, the detection of water vapor in the outer region of the disk, where comets are believed to form, and the estimated water mass of a few $\sim10^4-10^5$ Earth oceans, supports the scenario of impact delivery of water on terrestrial planets by means of icy bodies.




\begin{acknowledgements}
LP and CP acknowledges funding from the European 7$^\mathrm{th}$ Framework Program (FP7) (contract PIEF-GA-2009-253896, PERG06-GA-2009-256513), and from Agence Nationale pour la Recherche (ANR) (contract ANR-2010-JCJC-0504-01).
We also aknowledge funding from FP7-2011 (contract 284405), and the Service Commun de Calcul Intensif de l'IPAG for computations (contracts ANR-07-BLAN-0221, ANR-2010-JCJC-0504-01, ANR-2010-JCJC-0501-01). 
\end{acknowledgements}


\end{document}